\RequirePackage{fix-cm}
\documentclass[twocolumn]{svjour3}          % twocolumn
\smartqed  % flush right qed marks, e.g. at end of proof
\usepackage{graphicx}
\usepackage{algorithm}
\usepackage{algpseudocode}
\usepackage{amsmath}
\usepackage{microtype}
\usepackage{hyperref}
\usepackage{subcaption}
\usepackage{multirow}
\usepackage{titlesec}
\titleformat{\subsection}{\bfseries}{\thesubsection}{1em}{}
\usepackage{placeins}
\usepackage{makecell}

\begin{document}

\title{UNAD+: An Explainable Hybrid Framework for Unknown Network Attack Detection

}
% \subtitle{Do you have a subtitle?\\ If so, write it here}


% \author{Saif Alzubi         \and
%         Frederic Stahl %etc.
% }

% %\authorrunning{Short form of author list} % if too long for running head

% \institute{Saif Alzubi \at
%               Department of Computer Science, University of Exeter, Exeter, UK \\
%               \email{s.m.y.alzubi@exeter.ac.uk}           %  \\
% %             \emph{Present address:} of F. Author  %  if needed
%            \and
%            Frederic Stahl \at
%               German Research Center for Artificial Intelligence GmbH (DFKI), Marine Perception, Marie-Curie-Straße 1
% 26129 Oldenburg, Germany
% }

\author{Saif Alzubi \and Frederic Stahl}

\institute{
Saif Alzubi \at
Department of Computer Science, University of Exeter, \\Exeter, UK \\
\email{s.m.y.alzubi@exeter.ac.uk}
\and
Frederic Stahl \at
German Research Center for Artificial Intelligence GmbH (DFKI), Marine Perception, Marie-Curie-Straße 1, 26129 Oldenburg, Germany \\
\email{frederic.stahl@dfki.de}
}

\date{}

\maketitle

\begin{abstract}

The detection of previously unseen network attacks remains a major challenge for intrusion detection systems. Although supervised learning methods often perform well on known attack classes, they are limited when new attack types are not represented in the training data. Unsupervised methods are more suitable for detecting zero-day attacks, as they do not require labelled attack samples, but they often suffer from high false positive rates, which limits their real-world usefulness. This paper presents UNAD+, an enhanced framework for unknown network attack detection derived from the previously proposed Unknown Network Attack Detector (UNAD). UNAD+ combines a benign-only unsupervised ensemble with Weighted Majority Voting (WMV), a supervised refinement stage trained on pseudo-labelled detections, and a post hoc explainability layer that provides both local and global explanations. The framework was evaluated on the CICIDS2017 and NSL-KDD benchmark datasets. The results show that UNAD+ improves on the original UNAD framework, achieving F1-scores above 98\% across the benchmark datasets while significantly reducing false positives and enhancing transparency and deployment suitability through integrated explainability.

\keywords{Network Intrusion Detection  \and Machine Learning \and Explainable AI \and Zero-Day Attack Detection}
\end{abstract}

\section{Introduction}
\label{intro}

The increasing frequency and sophistication of network-based attacks pose significant challenges to modern cybersecurity infrastructures. Intrusion Detection Systems (IDSs) continue to serve as a core defensive mechanism by monitoring and analysing network traffic to identify potentially malicious activity \cite{chawla2018host}. As network environments grow in size and complexity, it becomes increasingly difficult to distinguish malicious traffic from legitimate behaviour, especially when attacks are designed to resemble normal activity \cite{11153648}. Despite substantial progress in machine learning-based IDS research, the reliable detection of unknown or previously unseen attacks remains a major challenge \cite{11173696}.

Many existing intelligent IDSs are based on supervised learning \cite{pinto2023survey}. These approaches often achieve strong performance when the attack classes encountered during deployment are well represented in the labelled training data \cite{ZOHOURIAN2024104034}. However, their effectiveness is limited to previously seen attack types, as they must be trained on labelled data that explicitly includes these attacks. In practice, newly emerging attacks, often referred to as zero-day attacks (i.e., attacks not previously seen by the system), may differ from the attack classes used during model training. As a result, supervised systems, although effective for known threats, often fail to generalise well to unfamiliar attack behaviours \cite{ahmad2022deep,zoppi2023algorithm}.

In contrast, unsupervised learning methods attempt to detect anomalous activity without relying on labelled attack data \cite{tong2024real}, making them well-suited for identifying previously unknown or zero-day attacks. Instead of learning from known attack signatures, these methods learn the characteristics of benign or normal traffic and treat substantial deviations from this profile as suspicious. This makes them particularly useful in settings where labelled examples of new threats are not yet available. However, unsupervised methods often suffer from high false positive rates (FPR) \cite{de2023unsupervised}, which limits their practical applicability and can increase the burden on security analysts. Therefore, although unsupervised methods are important for zero-day detection, their outputs often require further refinement before they can be used reliably in real-world environments.

These limitations motivate the development of hybrid IDS frameworks that combine the strengths of unsupervised and supervised learning. In such a design, an unsupervised stage can first identify suspicious traffic without requiring labelled attack data. In contrast, a supervised stage can learn from these detections and refine the final predictions. This reduces reliance on large volumes of labelled attack data while enabling the system to improve its detection capabilities after the initial discovery of a new threat. However, the design of such systems introduces further challenges, including how to combine the outputs of multiple anomaly detectors, how to reduce the effect of pseudo-labelling errors (i.e., errors arising when provisional benign or attack labels are assigned before the true class is confirmed), and how to make the final decisions understandable to human analysts.

In addition to detection performance, transparency is also an important requirement in intelligent IDSs. Although hybrid machine learning systems can improve detection performance, their internal reasoning is often difficult to interpret at the system level \cite{nolle2023explanations}. Even when some individual classifiers are explainable, the interaction among multiple models can still result in black-box behaviour. This lack of transparency may reduce analyst trust and complicate root-cause analysis, incident triage, and compliance in regulated sectors such as healthcare and finance \cite{hassija2024interpreting}. For this reason, explainable artificial intelligence (XAI) should be considered an integral part of deployable IDS design. By incorporating local and global explanation methods, the system can provide interpretable justifications for its decisions and support more transparent analysis \cite{gunning2019darpa}.

To address these issues, this paper presents UNAD+, an enhanced framework for unknown network attack detection that extends the previously proposed Unknown Network Attack Detector (UNAD) \cite{alzubi2021towards}. The framework preserves the benign-only unsupervised detection capability of the original model and introduces three main extensions. First, a Weighted Majority Voting (WMV) mechanism replaces simple majority voting so that stronger base detectors have greater influence on the final ensemble decision. Second, a supervised refinement stage is introduced, in which pseudo-labelled benign and attack flows generated by the unsupervised ensemble are used to train a secondary classifier that improves detection accuracy and reduces false positives. Third, a post hoc explainability layer is added, using local and global explanation methods to improve transparency and support analyst interpretation. In this way, UNAD+ combines unknown-attack detection, supervised refinement, and explainable decision support within a single modular framework for detecting zero-day or previously unseen attacks.

\sloppy {The proposed framework is evaluated on two widely used intrusion detection benchmark datasets, CICIDS2017 \cite{CICIDS2017} and NSL-KDD \cite{tavallaee2009detailed}. The empirical results show that UNAD+ significantly improves detection performance over the original UNAD baseline, while reducing false positives and enhancing transparency. Therefore, the contribution of this work is not limited to improving predictive performance; it also demonstrates how unsupervised detection, supervised refinement, and explainability can be integrated into a coherent IDS framework for more reliable zero-day attack detection.}

The paper is organised as follows: Section \ref{sec_related} reviews related work on ensemble-based intrusion detection, hybrid supervised-unsupervised approaches, and explainability techniques. Section \ref{sec_method} introduces the enhanced UNAD+ framework, detailing its weighted ensemble voting, supervised refinement using pseudo-labels, and post hoc explainability components. Section \ref{sec_Evaluation} presents the experimental evaluation. Section \ref{sec_Future} outlines future work, followed by conclusions in Section \ref{sec_conclusions}.

\section{Related Work}
\label{sec_related}

Machine learning has become an important component of modern intrusion detection systems (IDSs), and many existing approaches are based on supervised models \cite{parhizkari2023anomaly}. These methods often achieve strong classification performance when trained and tested on well-labelled datasets. However, their success depends heavily on the availability of labelled attack data \cite{hou2022handling}, which is often incomplete and biased toward previously known threats. This limits their usefulness in detecting zero-day or previously unseen attacks. In contrast, unsupervised anomaly detection methods aim to identify deviations from normal behaviour without relying on labelled attack samples \cite{nisioti2018intrusion}, making them more suitable for detecting unknown threats. Even so, their practical use remains challenging as they often produce high false positive rates \cite{qiu2023unraveling}, particularly in environments where benign traffic patterns are diverse or change over time \cite{11131124}. Techniques such as clustering, autoencoders, and density-based methods, including Isolation Forest (iForest) \cite{liu2012isolation} and Local Outlier Factor (LOF) \cite{breunig2000lof}, have been applied for this purpose, but their effectiveness varies across datasets and operational conditions.

Ensemble-based intrusion detection has been studied to improve robustness and mitigate the limitations of individual anomaly detectors. The aggregation of multiple base learners often leads to more stable and accurate results than standalone models \cite{sagi2018ensemble}. Several recent IDS studies aggregate ensemble outputs using majority voting \cite{ibrahim2025optimized,srivastava2025arlhnids,cheng2026ensemble}. For instance, this approach has been used to combine heterogeneous classifiers such as LSTM, KNN, and logistic regression \cite{ibrahim2025optimized}, ensembles of tree-based models \cite{srivastava2025arlhnids}, and IoT-focused intrusion detection frameworks \cite{cheng2026ensemble}. Although this approach is straightforward, it assumes that all base learners are equally reliable, an assumption that may not hold in practice. Some studies have explored weighting strategies based on model performance or confidence scores \cite{awad2025enhanced,alhowaide2021ensemble}, showing that stronger classifiers can contribute more effectively to decisions in ambiguous cases. Nevertheless, weighted voting remains relatively underused in the IDS literature, particularly in systems specifically designed for unknown attack detection.

Another line of research has focused on hybrid frameworks that combine unsupervised detection with supervised learning. The main motivation behind these systems is that an unsupervised stage can first detect suspicious or previously unknown traffic. In contrast, a supervised stage can then learn from those detections and improve the final classification results. Zoppi and Ceccarelli \cite{zoppi2021prepare}, for example, proposed a stacking-based framework in which anomaly outputs are refined by a supervised meta-classifier. Similarly, Kale et al. \cite{kale2022hybrid} combined clustering, semi-supervised learning, and supervised classification in a hybrid IDS setting. These studies demonstrate the value of combining different learning paradigms. However, many hybrid IDSs still provide limited detail on how pseudo-labels are managed, how error propagation is controlled, or how refinement is performed in a structured and repeatable way. As a result, although hybrid learning is promising, its integration into zero-day intrusion detection remains incomplete in many existing systems.

Explainability has also become an important issue as IDS architectures grow more complex. Explainable artificial intelligence (XAI) techniques such as Local Interpretable Model-Agnostic Explanations (LIME) \cite{ribeiro2016should} and SHapley Additive exPlanations (SHAP) \cite{lundberg2017unified} have been used to provide local explanations for machine learning predictions. In cybersecurity, such methods are valuable as they help analysts understand why a flow has been classified as malicious and which features influenced that decision. However, much of the existing work on XAI in security primarily provides explanations after detection, rather than embedding them as an explicit architectural component of the IDS workflow \cite{gaspar2024explainable,arreche2024xai}. Gaspar et al. \cite{gaspar2024explainable} apply LIME and SHAP to explain the outputs of a black-box IDS model. In contrast, Arreche et al. \cite{arreche2024xai} evaluate black-box XAI methods for network intrusion detection rather than incorporating explanation as an explicit component of the detection workflow, where it can support more transparent and actionable analysis.

A further difficulty in evaluating IDS frameworks lies in the datasets used for benchmarking. CICIDS2017 \cite{CICIDS2017} and NSL-KDD \cite{tavallaee2009detailed} are among the most widely used datasets in intrusion detection research, but they differ considerably in their traffic composition, attack diversity, and complexity. As a result, strong performance on one dataset does not necessarily indicate equivalent robustness on another. Although many studies report encouraging results on these benchmarks \cite{hnamte2023novel,de2023unsupervised}, fewer provide a combined solution that addresses the following three issues: detecting unknown attacks, refining detections through supervised learning, and providing transparent explanations of system decisions. However, limited attention has been given to systems that combine benign-only unsupervised detection, structured supervised refinement, and integrated explanation within a single framework. This gap motivates the UNAD+ framework proposed in this paper.

\section{UNAD+: Framework Design and Components}
\label{sec_method}

Building on the original UNAD framework, UNAD+ introduces three main extensions: weighted voting for improved ensemble decision-making, supervised refinement for more accurate and robust detection, and explainability for greater transparency and analyst support.

\subsection{The Original UNAD Framework}
\label{sec:2}

The original UNAD \cite{alzubi2021towards} is an unsupervised, ensemble-based intrusion detection framework designed to identify previously unseen attacks. In contrast to signature-based and supervised IDSs, UNAD is trained only on benign traffic. This enables the system to detect anomalous behaviour without requiring prior knowledge of attack signatures or labelled malicious instances. 

UNAD uses a heterogeneous ensemble of 100 base learners, consisting of 50 Local Outlier Factor (LOF) models \cite{breunig2000lof} and 50 Isolation Forest (iForest) models \cite{liu2012isolation}. Each learner is trained on a bootstrapped subset of benign data using bagging in order to promote diversity among the base detectors and reduce overfitting. The predictions produced by the ensemble are then combined using simple majority voting, where each model contributes one vote for either the benign or attack class, and the majority of votes determines the final prediction.

Although the original UNAD demonstrated that benign-only anomaly detection can identify unknown attacks, it also exhibited several limitations. First, the voting process assigned equal importance to all base learners, even though their performance could vary. Second, tying cases at the individual-prediction level required human intervention, reducing the level of automation in the detection process. Third, the framework lacked a mechanism to explain its decisions, limiting its suitability for practical deployment in settings where transparency and analyst trust are important.

\subsection{\textbf{Overview of UNAD+}}

\begin{figure*}[!t]
\centering
\includegraphics[width=0.95\linewidth]{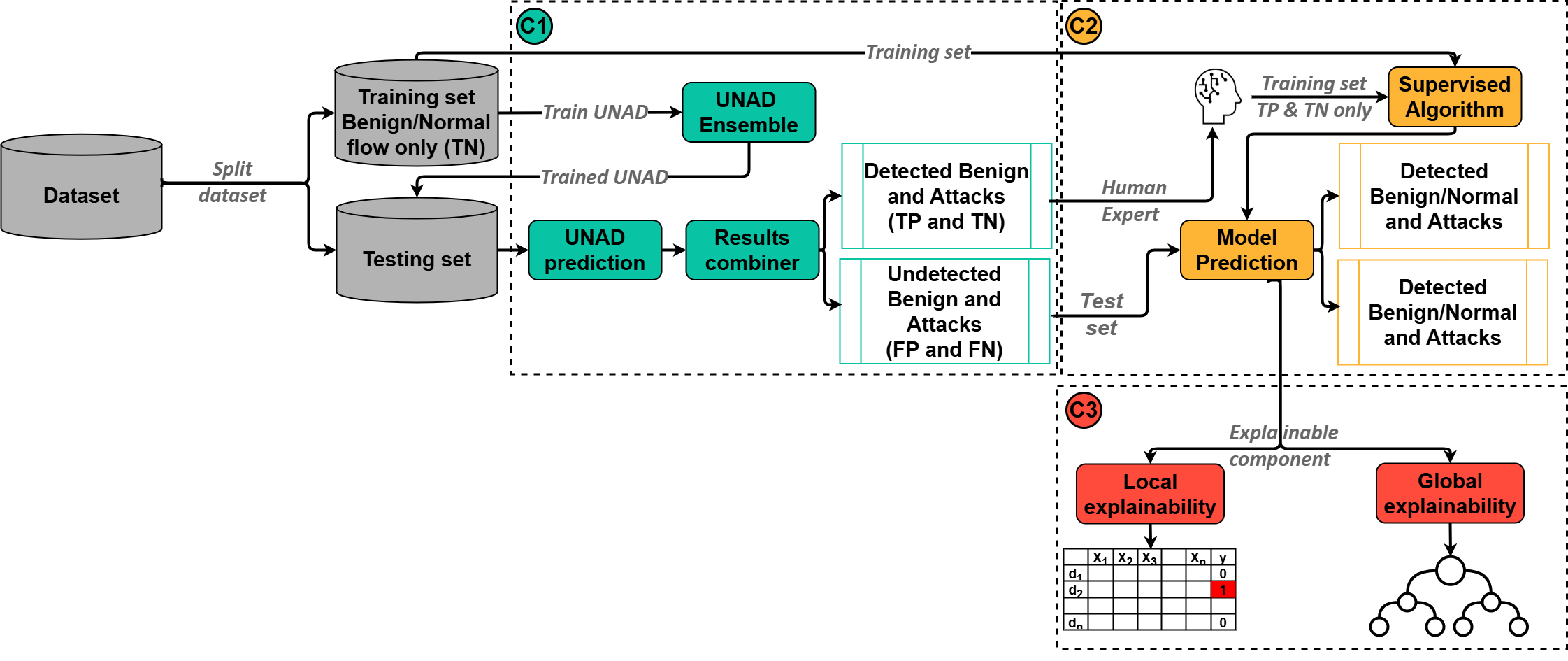}
\caption{Architecture of UNAD+: (1) unsupervised ensemble trained on benign data; (2) supervised refinement using pseudo-labels; (3) post hoc explainability for local and global interpretation.}
\label{Picture1}
\end{figure*}

To address the limitations of the original UNAD, this paper proposes UNAD+, which preserves the benign-only unsupervised detection capability of the original framework while introducing three main enhancements: (1) Weighted Majority Voting, (2) a supervised refinement stage, and (3) a post hoc explainability layer. Figure \ref{Picture1} shows the overall architecture of UNAD+, including the unsupervised ensemble (C1), the supervised refinement stage (C2), the human-in-the-loop checkpoint, and the explainability component (C3).

In this architecture, the first component detects suspicious or previously unseen traffic using a benign-only unsupervised ensemble. The second component refines these preliminary detections through supervised learning on pseudo-labelled outputs from the first stage, thereby improving classification quality and reducing false positives. The third component provides local and global explanations for the final decisions, improving transparency and helping analysts understand why a flow was classified as benign or malicious.

% \FloatBarrier
% \vspace{-0.9em}

To further improve robustness, a human-in-the-loop checkpoint is placed between the unsupervised ensemble and the supervised refinement stage \cite{Alzubi2}. This step allows a domain expert to review and validate a subset of pseudo-labelled detections before they are used for supervised learning. In this way, the framework reduces the risk of passing incorrect or unreliable labels from the unsupervised ensemble into the refinement component.

\subsection{\textbf{Ensemble Design and Bagging Strategy}} 

\begin{figure}[!t]
\centering
\includegraphics[width=1.0\linewidth]{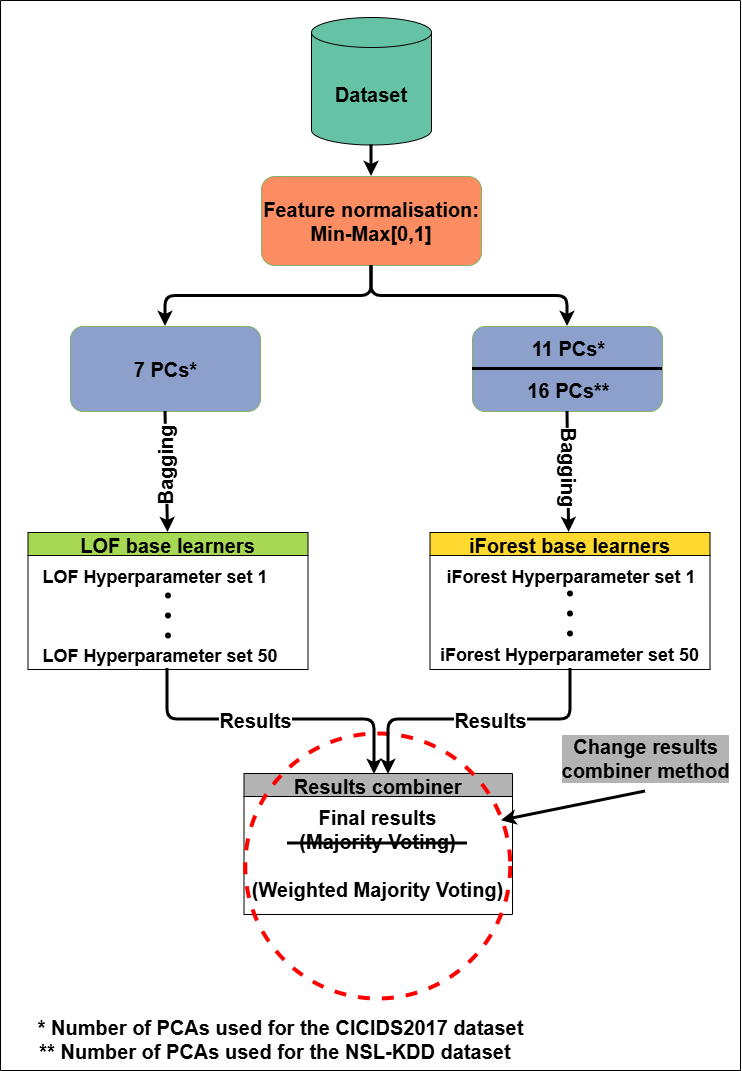}
\caption{Workflow of the unsupervised ensemble stage in UNAD+.}
\label{Picture2}
\end{figure}

The first stage of UNAD+ consists of data preprocessing followed by unsupervised ensemble construction. Figure \ref{Picture2} illustrates the structure of the unsupervised ensemble used in UNAD+. Before training, Principal Component Analysis (PCA) is applied to reduce dimensionality \cite{abdi2010principal}. For each dataset and detector, the number of retained principal components is determined through experimental tuning and selected based on the highest F1-score achieved \cite{Alzubi2}. The resulting reduced representations are then used as input to the unsupervised ensemble.

As in the original UNAD, the unsupervised ensemble combines 50 LOF learners and 50 iForest learners. This configuration was retained because an ensemble of 100 base learners produced the strongest overall performance, while the equal split between LOF and iForest avoided bias towards one learner type \cite{alzubi2021towards}. Among the anomaly detection methods initially considered, including Local Outlier Factor (LOF), Isolation Forest (iForest), Elliptic Envelope (EE), and One-Class Support Vector Machine (OCSVM), LOF and iForest achieved the strongest overall F1-scores and were therefore selected. Bagging is used to introduce variation across the ensemble by training each base learner on a different bootstrapped sample of the benign training data \cite{breiman1996bagging}. This improves generalisability and reduces the effect of instability in individual learners.

In addition to sample-level variation, further heterogeneity is introduced through the use of different high-performing hyperparameter combinations for each learner type. As a result, the ensemble benefits from both algorithmic heterogeneity, through the combined use of LOF and iForest, and internal variation, through multiple parameter settings and bootstrapped training subsets. Such a structure enables the system to better capture different forms of anomalous behaviour in network traffic.

\subsection{\textbf{Voting Strategy: Weighted Majority Voting (WMV)}} 
\label{subsec_WMV}

A key limitation of the original UNAD was its reliance on simple majority voting, which treats all base learners equally. In practice, however, not all base learners contribute equally to the final decision. Giving equal influence to weaker and stronger learners may reduce the effectiveness of the final decision and, in the case of an even number of learners, may also lead to tied votes that require tie-breaking.

To address this, UNAD+ replaces simple majority voting with Weighted Majority Voting (WMV). During validation, each base learner receives a weight proportional to its F1-score. At inference, predictions (0 for benign, 1 for attack) mask their weights. Scores for each class are summed, and the highest score is used as the ensemble output. 

This strategy gives stronger detectors greater influence on the final decision than weaker ones. As a result, the effect of unreliable models is reduced, and the ensemble is expected to yield more decisive predictions. Another practical advantage of WMV is that it removes the tie-related ambiguity observed in the original majority-voting version of UNAD, where tied votes were resolved in favour of the benign class, thereby introducing a bias towards benign predictions. Because the decision is based on continuous weighted scores rather than equal vote counts, exact ties become highly unlikely. WMV is therefore expected to improve the operational reliability of the framework while preserving the benefits of ensemble-based anomaly detection. The implementation of WMV in UNAD+ is shown in Algorithm~\ref{alg:wmv}.
\vspace{-0.6cm}

\begin{algorithm}[H]
\small
\caption{Weighted Majority Voting in UNAD+}
\label{alg:wmv}
\begin{algorithmic}[1]
\State \textbf{Input:}
\Statex \hspace{1.8em} Predictions $P = [p_1, p_2, \dots, p_n]$, where $p_i \in \{\text{benign}, \text{attack}\}$
\Statex \hspace{1.8em} F1-scores $F = [f_1, f_2, \dots, f_n]$ for each base learner
\State \textbf{Output:} Final class label for the instance
\State

\State Initialize: $Score_{benign} \gets 0$, $Score_{attack} \gets 0$

\For{$i = 1$ to $n$}
    \If{$p_i = \text{benign}$}
        \State $Score_{benign} \gets Score_{benign} + f_i$
    \Else
        \State $Score_{attack} \gets Score_{attack} + f_i$
    \EndIf
\EndFor

\If{$Score_{attack} > Score_{benign}$}
    \State \Return \text{attack}
\Else
    \State \Return \text{benign}
\EndIf

\end{algorithmic}
\end{algorithm}

\subsection{\textbf{Supervised Refinement}} 
\label{subsec_Refinement}

Although the weighted unsupervised ensemble provides the first stage of unknown attack detection, false positives and false negatives still occur. To improve the quality of these initial detections, UNAD+ introduces a supervised refinement stage after the WMV ensemble. In this stage, the correctly detected benign and attack instances produced by the unsupervised ensemble are used as pseudo-labels. These pseudo-labelled instances are combined with the original benign training data to create an expanded training set for supervised learning. This allows the second-stage classifier to learn more precise decision boundaries using traffic patterns that are closer to the operational detection setting than the original benign-only data alone.

Several supervised classifiers were evaluated for this role, including AdaBoost, Naive Bayes, K-Nearest Neighbours, and Random Forest. Before training, feature selection was applied using Information Gain (IG), and candidate feature subsets were assessed incrementally. Starting from the top five IG-ranked features, one additional feature was added at each iteration until the best 30 features had been evaluated. For each candidate subset, 10-fold cross-validation was used with a grid search to optimise model hyperparameters, and the final supervised classifier was selected based on the highest F1-score. To address class imbalance in the combined dataset, the Synthetic Minority Oversampling Technique (SMOTE) \cite{chawla2002smote} was applied after feature selection to balance the attack and benign classes to a 1:1 ratio.

Among the evaluated classifiers, Random Forest consistently achieved the strongest results and was therefore selected as the refinement classifier. In addition to its strong empirical performance, Random Forest is robust to overfitting and is also suitable for post hoc interpretability through feature importance analysis and surrogate modelling \cite{halabaku2024overfitting,aria2021comparison}. In this way, the second stage of UNAD+ serves as a refinement component that improves the quality of the initial anomaly detections and substantially reduces false positives.

\subsection{Explainability Layer} 

A further limitation of the original UNAD framework was the absence of interpretability. In practice, this limits the system's usefulness in environments where human analysts must inspect, validate, and trust model decisions. To address this issue, UNAD+ incorporates a post hoc explainability layer that supports both local and global interpretations.

For local explainability, Local Interpretable Model-Agnostic Explanations (LIME) \cite{ribeiro2016should} is applied to the Random Forest refinement classifier. LIME explains an individual prediction by perturbing the input instance, observing how the classifier’s output changes, and identifying the features that contribute most strongly to the final decision. LIME was selected for its model-agnostic nature, ease of application, and suitability for tabular network traffic data \cite{ribeiro2016should}. It also requires limited reconfiguration when the underlying model is updated. Compared with alternatives such as SHAP, which provides theoretically grounded Shapley-value explanations, \cite{aas2021explaining} LIME offers a more computationally efficient way to generate local explanations in this application setting \cite{ribeiro2016should}.

For global explainability, a Decision Tree surrogate model is trained to approximate the behaviour of the Random Forest. The surrogate is fitted using the input samples and the predictions generated by the Random Forest, and its fidelity is then evaluated to determine how well it reproduces the original model’s behaviour. Although the surrogate is simpler than the original classifier, it provides a more interpretable view of the overall decision logic, including dominant rules, feature interactions, and broad decision boundaries \cite{herbinger2023leveraging}. This global explanation helps analysts inspect the system's behaviour as a whole and supports auditing, validation, and model debugging.

Local and global explanations, therefore, serve complementary roles in UNAD+. Local explanations support the analysis of individual flows, while the surrogate model provides a broader understanding of how the refinement classifier behaves across the dataset.

\section{Experimental Evaluation and Analysis}
\label{sec_Evaluation}

The experimental evaluation assesses the detection performance, robustness, and explainability of UNAD+, with particular attention to the effects of weighted voting and supervised refinement compared with the original UNAD and baseline models.

\subsection{\textbf{Datasets}}

Two publicly available benchmark datasets were used to evaluate the proposed framework: CICIDS2017 \cite{CICIDS2017} and NSL-KDD \cite{tavallaee2009detailed}. CICIDS2017 was selected as it provides more recent and diverse attack scenarios in a realistic network setting, whereas NSL-KDD was included as a widely used benchmark with a different traffic structure and attack composition. Evaluating UNAD+ on both datasets therefore, supports a broader assessment of robustness and generalisability across different feature spaces, class distributions, and attack categories. Table \ref{tab:dataset_summary} summarises the main characteristics of the two datasets used in the evaluation.

CICIDS2017 is a publicly available benchmark dataset generated by the Canadian Institute for Cybersecurity over five days and consists of about 3 million data instances \cite{CICIDS2017}. Developed in a realistic network environment, it includes multiple operating systems and commonly used protocols such as HTTP, HTTPS, FTP, SSH, and email protocols. Compared with legacy IDS benchmarks, CICIDS2017 provides a more recent and more representative view of modern network traffic. The dataset covers 14 attacks grouped into seven categories: Brute Force, Heartbleed, Botnet, DoS, DDoS, Web Attacks, and Infiltration. The attack types considered in this study were DoS Hulk, Port Scan, DDoS, DoS GoldenEye, FTP-Patator, SSH-Patator, DoS Slowloris, DoS Slowhttptest, Botnet, Web Attack-Brute Force, Web Attack-XSS, Infiltration, Web Attack-SQL Injection, and Heartbleed. Traffic features were extracted from PCAP files using CICFlowMeter, which generates 84 network traffic features together with the class label.

NSL-KDD is an updated version of the KDD99 dataset, which was originally derived from the DARPA98 dataset. The benchmark was introduced to address known limitations of KDD99, particularly the large number of duplicate records that can bias model evaluation \cite{tavallaee2009detailed}. NSL-KDD consists of two files: KDDTrain+, containing 125,973 records, and KDDTest+, containing 22,544 records. Four main attack categories are represented in addition to normal traffic: Denial of Service (DoS), User to Root (U2R), Remote to Local (R2L), and Probe. Its 43 features are grouped into three types: basic features derived from TCP/IP connections, traffic features computed over time windows, and content features designed to capture suspicious behaviour relevant to classes such as R2L and U2R. Given its different structure and attack composition, NSL-KDD provides a useful complement to CICIDS2017 in evaluating the proposed framework across heterogeneous benchmark settings.

\begin{table*}[t]
\centering
\caption{Summary of the datasets used in the evaluation.}
\label{tab:dataset_summary}
\begin{tabular}{lrp{3.6cm}p{6.0cm}}
\hline\noalign{\smallskip}
\textbf{Dataset} & \textbf{Features } & \textbf{Total records} & \textbf{Attack categories} \\
\noalign{\smallskip}\hline\noalign{\smallskip}
CICIDS2017 & 84 & 2,829,463 &
\makecell[l]{Brute Force, Heartbleed, Botnet, DoS, DDoS,\\ Web Attacks, Infiltration} \\
NSL-KDD & 43 &
\makecell[l]{125,973 (KDDTrain+)\\ + 22,544 (KDDTest+)} &
DoS, U2R, R2L, Probe \\
\noalign{\smallskip}\hline
\end{tabular}
\end{table*}

\subsection{\textbf{Preprocessing and Hyperparameter Tuning}}

For CICIDS2017, preprocessing began with the removal of duplicate, missing, and invalid records. Features representing IDs, IP addresses, and ports were then discarded, as they do not provide stable discriminatory value for intrusion detection and may bias the model towards dataset-specific characteristics. The label field was converted to binary, with benign flows assigned 0 and all attack types assigned 1. The data were subsequently normalised using Min-Max scaling. After preprocessing, the dataset retained 76 final features and contained 2,827,672 flows, of which 2,271,117 were benign, and 556,555 were attacks. To reflect the zero-day detection setting, benign Monday traffic, comprising 529,445 flows, was used as the training set for the unsupervised ensemble. PCA was then applied for dimensionality reduction, with 7 principal components used for LOF and 11 for iForest.  

For NSL-KDD, preprocessing followed a similar procedure, including the removal of missing and duplicate records, binary label encoding, and Min-Max scaling. In addition, the categorical features protocol type, service, and flag were converted to numeric form using one-hot encoding, resulting in 122 final features. The KDDTrain+ file was first split into 60\% training and 40\% hold-out data. Since the unsupervised ensemble was trained only on benign traffic, all attack instances were removed from the training portion, leaving 40,405 normal records. The remaining 40\% of KDDTrain+ was then combined with KDDTest+ and re-split into a validation set of 37,791 instances and a test set of 35,140 instances using stratified random sampling so that both sets contained representative attack classes. PCA was again applied, with 7 principal components used for LOF and 16 for iForest.

Hyperparameters for both the unsupervised models and the supervised refinement classifier were optimised using grid search, with F1-score used as the primary selection criterion. For LOF, the contamination parameter was explored from 0.01 to 0.50 in steps of 0.01, and the number of neighbours from 5 to 50 in steps of 5. For iForest, contamination was explored over the same range. At the same time, the number of estimators was varied from 50 to 600 in steps of 50, and the maximum number of samples was assessed using the auto setting together with 25\%, 50\%, 75\%, and 100\% of the training data. For the supervised refinement stage, Random Forest was selected as the final classifier after tuning the candidate models described in Section \ref{subsec_Refinement}. The explored hyperparameter ranges for Random Forest included the number of estimators from 100 to 500 in steps of 50, the maximum depth over the default setting, together with values from 5 to 15 in steps of 5, the minimum samples split from 2 to 8 in steps of 2, and the minimum samples leaf over the default setting together with values from 2 to 6 in steps of 2. Table \ref{tab:hyperparameters} summarises the hyperparameter ranges explored and the final values selected for the models used in the experiments.

\begin{table*}[t]
\centering
\caption{Hyperparameter tuning ranges and selected values for LOF, iForest, and Random Forest.}
\label{tab:hyperparameters}
\scriptsize
\begin{tabular}{llp{4.8cm}p{2.7cm}p{2.5cm}}
\hline\noalign{\smallskip}
\textbf{Model} & \textbf{Parameter} & \textbf{Range explored} & \shortstack{\textbf{Final value} \\ \textbf{(CICIDS2017)}} & \shortstack{\textbf{Final value} \\ \textbf{(NSL-KDD)}} \\
\noalign{\smallskip}\hline\noalign{\smallskip}
\multirow{3}{*}{LOF}
& PCA components & 2--15 (CICIDS2017), 2--17 (NSL-KDD) & 7 & 7 \\
& contamination & 0.01--0.50 & 0.07 & 0.14 \\
& n\_neighbors & 5--50 & 30 & 5 \\
\noalign{\smallskip}
\multirow{4}{*}{iForest}
& PCA components & 2--15 (CICIDS2017), 2--17 (NSL-KDD) & 11 & 16 \\
& contamination & 0.01--0.50 & 0.24 & 0.10 \\
& n\_estimators & 50--600 & 400 & 100 \\
& max\_samples & auto, 25\%, 50\%, 75\%, 100\% & 25\% & 100\% \\
\noalign{\smallskip}
\multirow{4}{*}{Random Forest}
& n\_estimators & 100--500 & 100 & 300 \\
& max\_depth & Default, 5, 10, 15 & 10 & 15 \\
& min\_samples\_split & 2--8 & 8 & 4 \\
& min\_samples\_leaf & Default, 2--6 & 2 & Default \\
\noalign{\smallskip}\hline
\end{tabular}
\end{table*}

\subsection{\textbf{Evaluation Metrics}}

The evaluation uses standard binary classification measures, namely Precision, Recall, F1-score, ROC-AUC, and False Positive Rate (FPR). In all experiments, attack instances were treated as anomalous and benign or normal instances as non-anomalous. These measures were selected to assess detection quality and operational reliability jointly. In intrusion detection, it is important not only to detect malicious traffic, but also to limit the number of benign flows incorrectly classified as attacks.

Precision measures the proportion of detected attacks that are truly malicious, while Recall measures the proportion of actual attacks that are successfully detected. F1-score combines these two measures and was therefore used as the main performance indicator throughout the model-selection process. ROC-AUC was used to assess class separation across thresholds, while FPR quantified the extent to which benign traffic was incorrectly flagged as malicious.

\subsection{\textbf{Overall Performance on CICIDS2017 and NSL-KDD}}

Detection performance was evaluated for four settings: the original UNAD using simple majority voting, the weighted ensemble version (UNAD+ WMV Ensemble), the standalone supervised refinement classifier, and the full combined UNAD+ framework. The standalone supervised refinement classifier was evaluated separately on ensemble error cases, namely false positives and false negatives, and is included here as a diagnostic comparison. Tables~\ref{tab:cicids_transpose} and~\ref{tab:nslkdd_transpose} report the results obtained on CICIDS2017 and NSL-KDD, respectively, across the standard evaluation metrics. 

On CICIDS2017, the original UNAD achieved an F1-score of 75.19\%, while the UNAD+ WMV Ensemble achieved 74.91\%. The supervised refinement classifier achieved an F1-score of 90.59\%, and the final combined UNAD+ framework reached an F1-score of 98.31\%, along with 99.44\% precision, 99.21\% recall, and 98.52\% ROC-AUC. On NSL-KDD, the original UNAD achieved an F1-score of 93.38\%, while UNAD+ WMV Ensemble achieved 93.16\%. The supervised refinement classifier achieved an F1-score of 78.47\%, and the full UNAD+ framework reached 98.25\%, with 97.26\% precision, 99.26\% recall, and 98.24\% ROC-AUC.

\begin{table*}[t]
\centering
\caption{CICIDS2017 Results (in \%).}
\label{tab:cicids_transpose}
\begin{tabular}{lccccc}
\hline\noalign{\smallskip}
\textbf{Model} & \textbf{Accuracy} & \textbf{Precision} & \textbf{Recall} & \textbf{F1-score} & \textbf{ROC-AUC} \\
\noalign{\smallskip}\hline\noalign{\smallskip}
Original UNAD & 87.23 & 70.99 & 79.92 & 75.19 & 84.74 \\
UNAD+ WMV Ensemble & 86.84 & 69.57 & 81.14 & 74.91 & 84.90 \\
Standalone Supervised Refinement Classifier & 93.86 & 96.69 & 85.22 & 90.59 & 91.83 \\
\textbf{Final Combined Framework} & \textbf{99.19} & \textbf{99.44} & \textbf{99.21} & \textbf{98.31} & \textbf{98.52} \\
\noalign{\smallskip}\hline
\end{tabular}
\end{table*}

\begin{table*}[t]
\centering
\caption{NSL-KDD Results (in \%).}
\label{tab:nslkdd_transpose}
\begin{tabular}{lccccc}
\hline\noalign{\smallskip}
\textbf{Model} & \textbf{Accuracy} & \textbf{Precision} & \textbf{Recall} & \textbf{F1-score} & \textbf{ROC-AUC} \\
\noalign{\smallskip}\hline\noalign{\smallskip}
Original UNAD & 93.45 & 93.90 & 92.86 & 93.38 & 93.44 \\
UNAD+ WMV Ensemble & 93.22 & 93.52 & 92.80 & 93.16 & 93.22 \\
Standalone Supervised Refinement Classifier & 74.02 & 69.76 & 89.67 & 78.47 & 73.10 \\
\textbf{Final Combined Framework} & \textbf{98.24} & \textbf{97.26} & \textbf{99.26} & \textbf{98.25} & \textbf{98.24} \\
\noalign{\smallskip}\hline
\end{tabular}
\end{table*}

These results show that the weighted voting stage alone did not improve the F1-score relative to the original UNAD. However, the higher F1-scores obtained by the original UNAD should not be interpreted as evidence of superior first-stage performance, since under simple majority voting, tied votes were resolved in favour of the benign class, thereby introducing a systematic bias towards benign predictions.

Overall, the strongest performance was achieved by the full UNAD+ framework after combining weighted ensemble detection with supervised refinement. On CICIDS2017, the F1-score increased from 75.19\% in the original UNAD to 98.31\% in the final combined framework, while on NSL-KDD, the F1-score increased from 93.38\% to 98.25\%. These results confirm that the main contribution of UNAD+ lies not in the weighted voting stage in isolation, but in the combination of weighted ensemble detection and supervised refinement using pseudo-labelled data.

\subsection{\textbf{Impact of Supervised Refinement}}

\begin{figure}[H]
\centering
\includegraphics[width=1.0\linewidth]{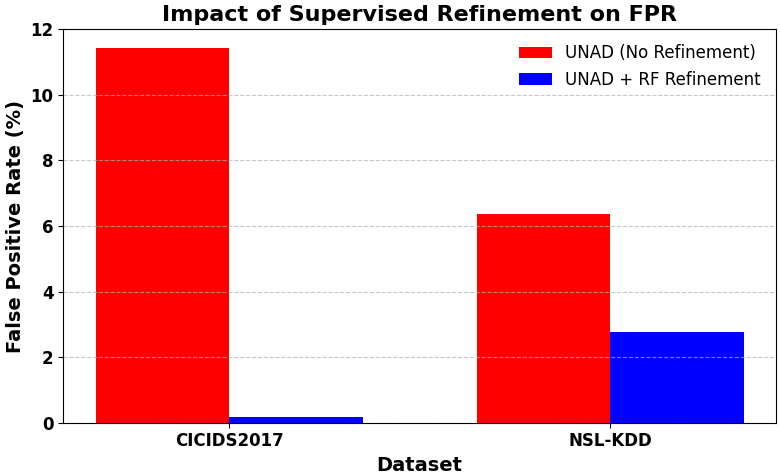}
\caption{Impact of supervised refinement on false positive rate (FPR) for CICIDS2017 and NSL-KDD.}
\label{Picture3}
\end{figure}

To assess the contribution of the second stage of the framework, an ablation analysis was conducted by comparing the weighted ensemble alone with the framework after supervised refinement. The main objective of this analysis was to examine the effect of the refinement stage on false-positive reduction, since high false-positive rates remain a major limitation of anomaly-based intrusion detection systems. Figure \ref{Picture3} presents the false positive rate (FPR) before and after applying the supervised refinement stage on CICIDS2017 and NSL-KDD.

The supervised refinement stage significantly reduced the FPR on both datasets. On CICIDS2017, the FPR decreased from 11.34\% in the weighted ensemble to 0.18\% after refinement, representing a reduction of more than 98\%. On NSL-KDD, the FPR decreased from 6.37\% to 2.77\%, corresponding to a reduction of more than half. These results show that the refinement stage effectively suppressed a large proportion of the spurious alerts produced by the unsupervised ensemble.

This improvement reflects the role of the second-stage classifier. After the weighted ensemble identifies suspicious traffic, its correctly detected benign and attack instances are used as pseudo-labelled data for supervised learning. This enables the refinement classifier to learn more precise decision boundaries from samples that are closer to the operational detection setting than the original benign-only training data. As a result, the second stage can correct a large proportion of the residual errors produced by the first stage, particularly when benign and malicious behaviours overlap.

The supervised refinement classifier was evaluated on dedicated test sets consisting of instances previously misclassified by the ensemble, namely false positives and false negatives. This corresponded to 151,245 instances for CICIDS2017 and 2,383 instances for NSL-KDD. The strong reduction in FPR therefore indicates that the second stage does not merely repeat the decisions of the ensemble, but instead provides a meaningful correction mechanism that improves the reliability of the framework.

The results confirm that the supervised refinement stage plays an important role in improving the real-world usefulness of UNAD+. It addresses one of the main weaknesses of unsupervised anomaly detection by significantly reducing false positives while preserving strong detection performance, thereby directly contributing to the improvement achieved by the final combined framework.

\subsection{\textbf{Effect of Ensemble Voting Strategy}}

\begin{figure}
\centering
\includegraphics[width=1.025\linewidth]{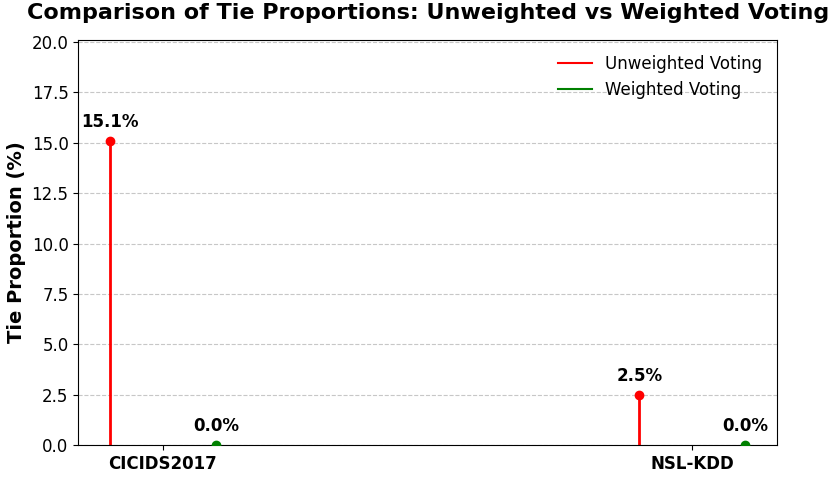}
\caption{Comparison of tie-case proportions under majority voting and weighted majority voting on CICIDS2017 and NSL-KDD.}
\label{Picture4}
\end{figure}

To evaluate the effect of the ensemble voting strategy, Weighted Majority Voting (WMV) was compared with the original simple majority voting approach on both the CICIDS2017 and NSL-KDD datasets. Figure \ref{Picture4} presents the proportion of tie cases produced by the two voting strategies on the two datasets. Under simple majority voting, equal numbers of benign and attack votes resulted in tied cases, which were resolved in favour of the benign class label. This introduced ambiguity into the first stage of detection and reduced the operational reliability of the original UNAD framework.

Simple majority voting resulted in ties in 15.1\% of predictions on CICIDS2017 and 2.5\% on NSL-KDD. In contrast, WMV reduced ties to 0\% across both datasets by assigning each base learner a weight according to its F1-score, thereby giving greater influence to stronger detectors. This made the first-stage ensemble more decisive by eliminating tied votes, which under simple majority voting were resolved in favour of the benign class. The effect was more pronounced on CICIDS2017, where the diversity of traffic types appears to have led to greater variation in the predictions of the base learners.

Although the overall metric differences between majority voting and WMV were relatively small, WMV offered a significant practical advantage. By eliminating tie cases, it removed one of the main limitations of the original UNAD and improved the consistency and operational suitability of the ensemble, since the first-stage detector no longer depended on benign-default resolution when equal votes occurred.

\subsection{\textbf{Class-Level Detection Analysis}}

\begin{figure*}[t]
\centering
\includegraphics[width=1.0\linewidth]{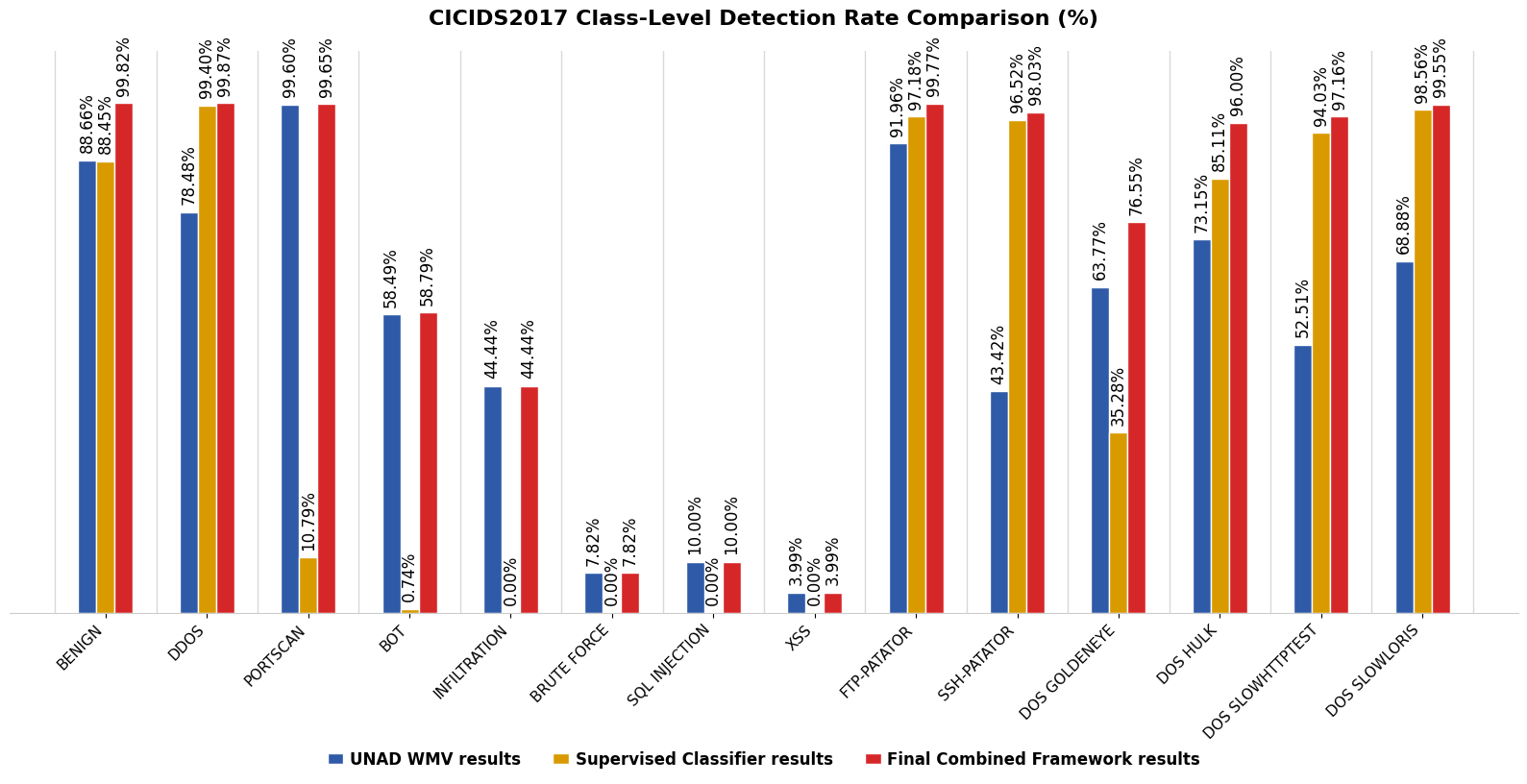}
\caption{Comparison of class-level detection rates on CICIDS2017 for the weighted ensemble (UNAD+ WMV), the supervised refinement classifier, and the final combined framework (in \%).}
\label{Picture5}
\end{figure*}

\begin{figure*}[t]
\centering
\includegraphics[width=0.70\linewidth]{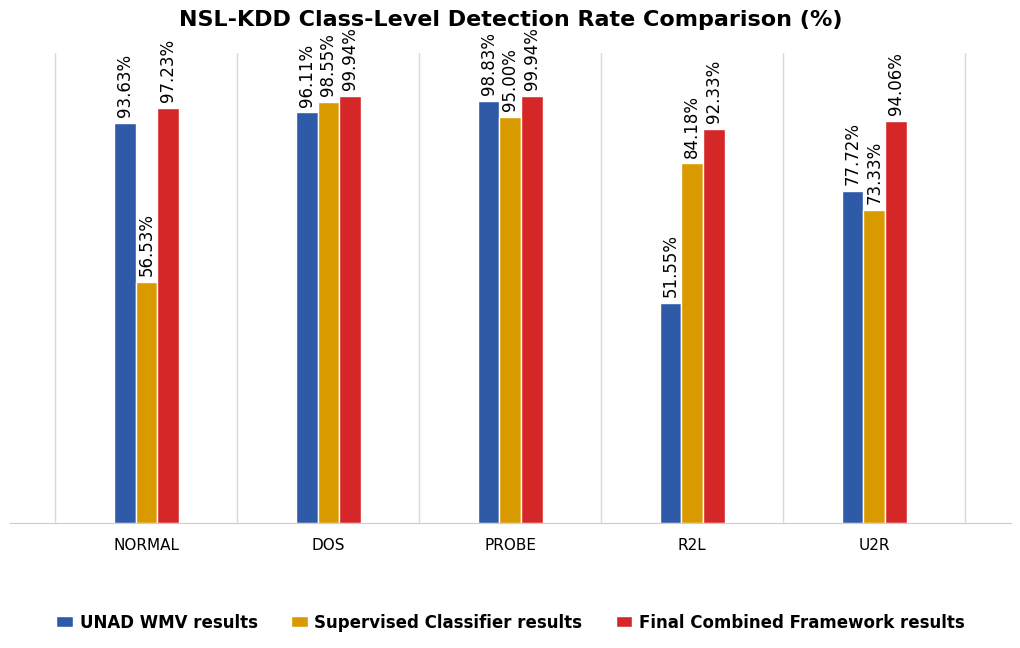}
\caption{Comparison of class-level detection rates on NSL-KDD for the weighted ensemble (UNAD+ WMV), the supervised refinement classifier, and the final combined framework (in \%).}
\label{Picture6}
\end{figure*}

To assess the behaviour of the framework beyond aggregate metrics, class-level detection rates were analysed for both datasets. Figures \ref{Picture5} and \ref{Picture6} compare three stages of the proposed system: the weighted unsupervised ensemble, the supervised refinement classifier evaluated separately, and the final combined framework, in which the weighted ensemble is followed by the supervised refinement stage, on CICIDS2017 and NSL-KDD, respectively. This analysis highlights which traffic classes benefited most from the second stage and which remained difficult to detect.

On CICIDS2017, the supervised refinement stage contributed significantly to improving the detection of several attack types. As shown in Figure \ref{Picture5}, the final combined framework improved the detection rates for benign traffic and most attack classes compared with the weighted ensemble alone. The benign detection rate increased by just over 11\%, from 88.66\% to 99.82\%. Among the attack classes, the largest improvements were observed for SSH-Patator, which increased from 43.42\% to 98.03\%, and DoS Slowhttptest, which improved to 97.71\%. Strong gains were also observed for DoS Slowloris, DDoS, and DoS Hulk, while FTP-Patator and DoS GoldenEye also improved after the second-stage refinement. In contrast, port scans and Bot attacks improved only marginally.

These results indicate that the second stage was particularly effective for attack categories that were initially detected with moderate performance by the ensemble, but for which sufficient pseudo-labelled training instances were available. In contrast, the classes that did not improve were those for which only very small numbers of pseudo-labelled examples reached the supervised stage. As a result, the refinement classifier was unable to improve detection for web attacks and infiltration.

The weak performance on these rare CICIDS2017 classes is attributable to the very small number of pseudo-labelled instances that reached the supervised refinement stage. For example, the supervised model was trained on only 59 brute force, 13 XSS, 8 infiltration, and 1 SQL injection instances. In addition, dimensionality reduction through PCA may also have reduced information useful for distinguishing these rare attack types.

NSL-KDD showed a similar pattern, but with a more consistent improvement across classes. As shown in Figure \ref{Picture6}, the final combined framework improved the detection rate for normal traffic and all attack categories after adding the supervised refinement classifier. The largest gain was observed for R2L attacks, where the detection rate increased from 51.55\% to 92.33\%. U2R attacks also improved substantially, from 77.72\% to 94.06\%. In addition, normal and DoS traffic improved by more than 3\%, while Probe attacks improved by about 1\%. These results suggest that the refinement stage was especially useful for rare and difficult attack classes in NSL-KDD, while also preserving the already strong performance of the weighted ensemble on the easier categories.

The class-level analysis shows that the main benefit of the hybrid framework lies in its ability to improve classes that are either highly variable or initially difficult for the unsupervised ensemble alone. At the same time, it shows that the success of the refinement stage depends on the availability of sufficiently informative pseudo-labelled examples. Therefore, while UNAD+ improves detection across most major traffic classes and significantly strengthens several difficult categories, its performance remains limited for very rare classes that are weakly represented in the first-stage detections.

\subsection{\textbf{Evaluation of the Explainability Component}}

\begin{figure*}[t]
    \centering
    \begin{subfigure}[t]{0.486\textwidth}
        \centering
        \includegraphics[width=\linewidth]{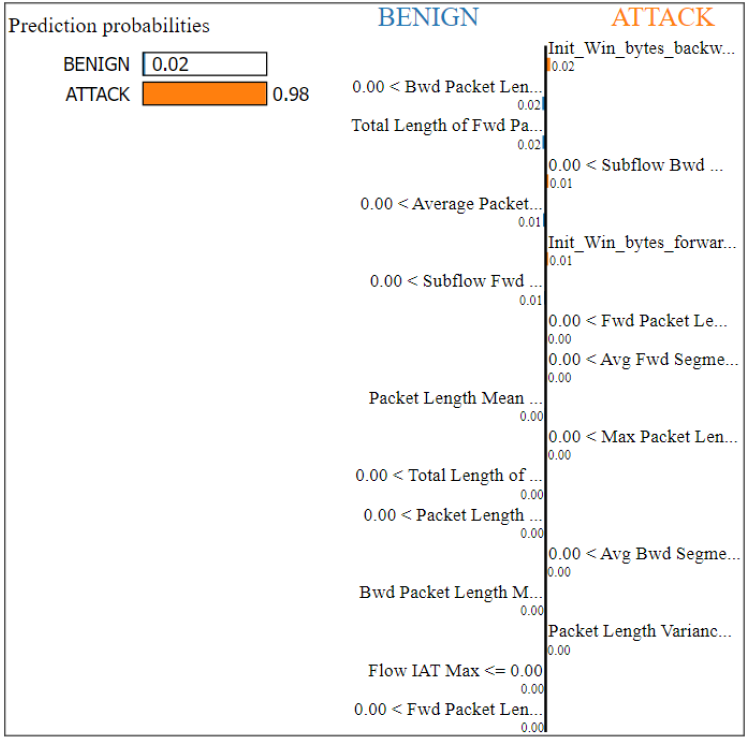}
        \caption{LIME explanation for a correctly classified attack instance.}
        \label{fig:7a}
    \end{subfigure}
    \hfill
    \begin{subfigure}[t]{0.48\textwidth}
        \centering
        \includegraphics[width=\linewidth]{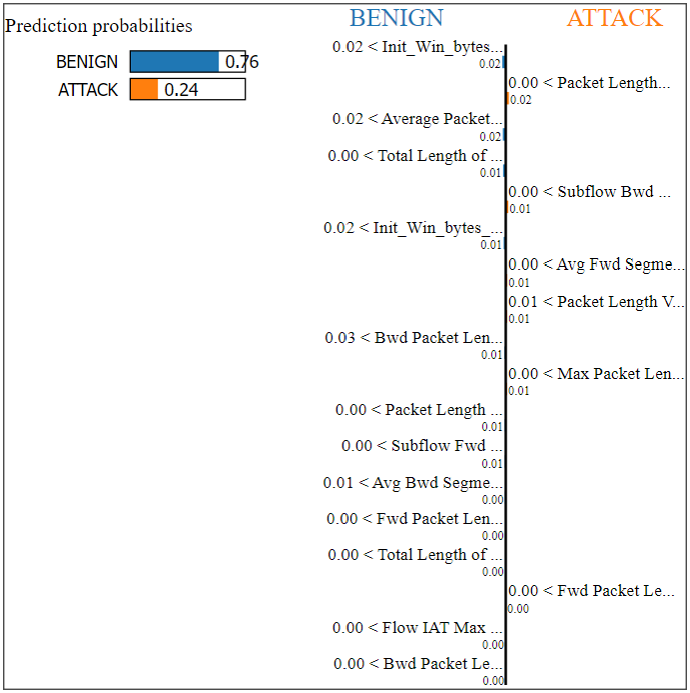}
        \caption{LIME explanation for an incorrectly classified attack instance.}
        \label{fig:7b}
    \end{subfigure}
    \caption{Local LIME explanations for attack instances on CICIDS2017.}
    \label{fig:cicids2017attacklevel}
\end{figure*}

\begin{figure*}[t]
    \centering
    \begin{subfigure}[t]{0.486\textwidth}
        \centering
        \includegraphics[width=\linewidth]{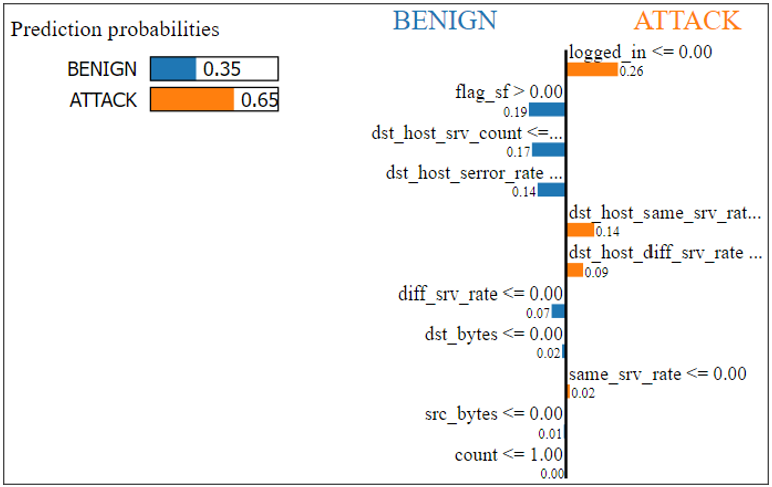}
        \caption{LIME explanation for a correctly classified attack instance.}
        \label{fig:8a}
    \end{subfigure}
    \hfill
    \begin{subfigure}[t]{0.48\textwidth}
        \centering
        \includegraphics[width=\linewidth]{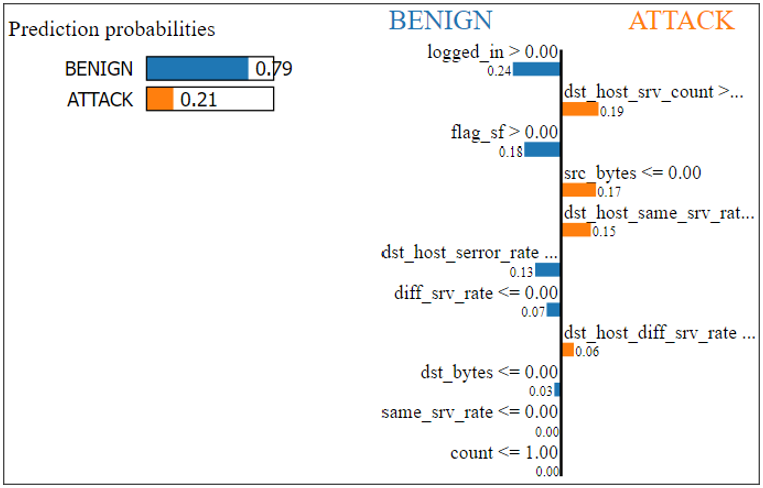}
        \caption{LIME explanation for an incorrectly classified attack instance.}
        \label{fig:8b}
    \end{subfigure}
    \caption{Local LIME explanations for attack instances on NSL-KDD.}
    \label{fig:nslkddattacklevel}
\end{figure*}

\begin{figure*}[t]
    \centering
    \begin{subfigure}[t]{1.0\textwidth}
        \centering
        \includegraphics[width=\linewidth]{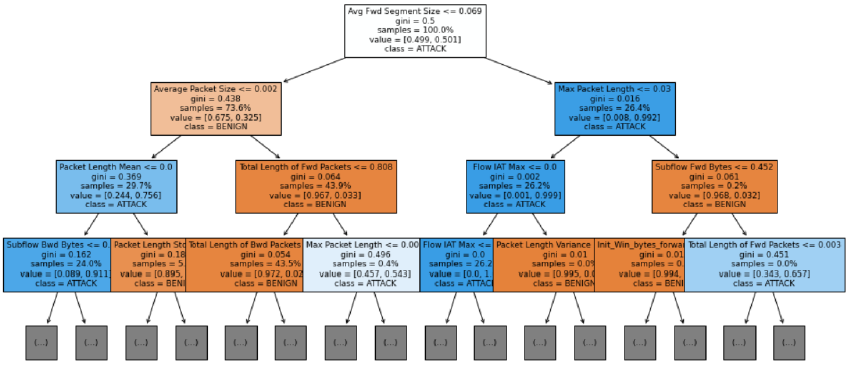}
        \caption{Decision Tree surrogate for CICIDS2017.}
        \label{fig:9a}
    \end{subfigure}

    \vspace{0.5cm}

    \begin{subfigure}[t]{1.0\textwidth}
        \centering
        \includegraphics[width=\linewidth]{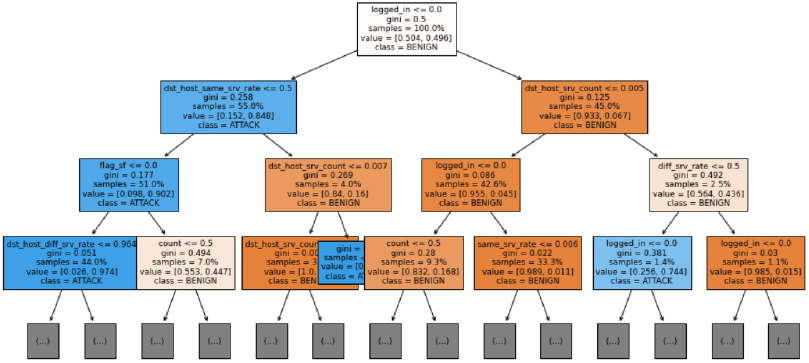}
        \caption{Decision Tree surrogate for NSL-KDD.}
        \label{fig:8b}
    \end{subfigure}
    \caption{Decision Tree surrogate models used for global explanation of the refinement classifier on CICIDS2017 and NSL-KDD.}
    \label{fig:globalattacklevel}
\end{figure*}

The explainability component was examined in terms of plausibility, qualitative stability, and fidelity. Plausibility refers to whether the explanations highlight traffic features that are meaningful for the predicted class. Stability refers here to whether local explanations remain broadly consistent across repeated inspection and small input changes. Fidelity refers to how accurately a simpler surrogate model can reproduce the behaviour of the Random Forest refinement classifier. Figure \ref{fig:cicids2017attacklevel} presents representative local explanations generated by LIME for CICIDS2017, Figure \ref{fig:nslkddattacklevel} presents the corresponding local explanations for NSL-KDD, and Figure \ref{fig:globalattacklevel} shows the Decision Tree surrogate models used for global interpretation.

In terms of plausibility, the local explanations indicated that the refinement classifier relied on coherent and interpretable traffic characteristics. On CICIDS2017, LIME repeatedly highlighted features such as Init Win Bytes Backwards, Subflow Bwd Packets, Fwd Packet Length Max, and Packet Length Variance for correctly classified attack instances, including SSH-Patator and DoS Slowhttptest. As shown in Figure\hyperref[fig:7a]{~\ref*{fig:cicids2017attacklevel}\subref*{fig:7a}}, these features appear consistently across representative attack cases rather than as isolated patterns. This supports the interpretation that the classifier’s local reasoning aligns with meaningful indicators of abnormal flow behaviour rather than spurious correlations.

A similar pattern was observed on NSL-KDD. For representative DoS and R2L instances, the most influential LIME features included dst bytes, flag, and service. These are interpretable network-level attributes, and their repeated appearance in the local explanations suggests that the refinement classifier is using features that are relevant to suspicious traffic behaviour. As illustrated in Figure \hyperref[fig:8a]{~\ref*{fig:nslkddattacklevel}\subref*{fig:8a}}, the usefulness of the local explanation layer lies not only in producing explanations, but also in grounding those explanations in features that an analyst can interpret and assess.

With respect to stability, the local explanations appeared broadly consistent across repeated inspection of representative cases and were not highly sensitive to small input changes. This is relevant because an explanation method has limited practical value if minor changes in the input produce substantially different explanations. In the present case, the observed consistency suggests that LIME is suitable as an analyst-facing explanation tool within the framework.

The local explanation layer was also informative in analysing classifier errors. As shown in Figures \hyperref[fig:7b]{~\ref*{fig:cicids2017attacklevel}\subref*{fig:7b}} and \hyperref[fig:8b]{~\ref*{fig:nslkddattacklevel}\subref*{fig:8b}}, incorrectly classified attack instances on both datasets still yielded interpretable feature-level explanations, making it possible to identify which features pushed the prediction towards the wrong class. This is analytically important, as it shows that the explanation layer remains informative even when the classifier fails. In such cases, the issue is not the absence of interpretable structure, but rather that the balance of influential features favours the incorrect class. This makes the explanations useful for debugging and for identifying systematic weaknesses in the refinement classifier.

Regarding fidelity, the global explainability layer was evaluated by training a Decision Tree surrogate model to approximate the predictions of the Random Forest refinement classifier. Fidelity was measured using accuracy against the Random Forest outputs, and the resulting score exceeded 99\% on both datasets. As shown in Figure \ref{fig:globalattacklevel}, the surrogate models capture the dominant decision rules and feature splits of the refinement classifier while remaining substantially easier to inspect than the original model. This indicates that the broader behaviour of the second-stage classifier can be approximated in an interpretable form while preserving its main decision rules, feature splits, and overall classification logic.

The global explainability layer was further supported by a rule-extraction component derived from the surrogate Decision Tree. This complements the surrogate models shown in Figure \ref{fig:globalattacklevel} by providing a more structured view of the classifier’s broader decision logic, including the dominant feature thresholds and decision paths associated with attack and benign predictions. In this way, the global explanation layer supports auditing, validation, and the identification of possible operational blind spots, for example, cases where the classifier relies too heavily on a small number of features or fails to distinguish rare attack patterns effectively.

Overall, the explainability evaluation suggests that the post hoc explanation layer improves the interpretability of the framework. The local explanations support plausibility by highlighting meaningful traffic features and showing broad consistency across representative analyses, while the global surrogate maintains high fidelity to the refinement classifier. This combination improves transparency in ways that support analyst trust, auditing, debugging, and security review.

\section{Future Work}
\label{sec_Future}

One important challenge not addressed in the present study concerns encrypted traffic. In contemporary network environments, encrypted communication is increasingly prevalent, which makes intrusion detection more difficult because visibility into packet content is reduced and detection must rely more heavily on flow-level and metadata-based features. Handling encrypted attacks was outside the scope of the work presented here, which focused on evaluating UNAD+ on benchmark datasets under the current feature space. Future work will therefore examine whether stronger flow-level, timing-based, and temporal descriptors can improve detection performance under encrypted traffic conditions, particularly in reducing false positives, while preserving the framework's interpretability.

\section{Conclusions}
\label{sec_conclusions}

This paper presented UNAD+, an enhanced intrusion detection framework designed to address three limitations of the original UNAD: the reliance on simple majority voting, the high false positive rate, and the lack of interpretability. The proposed framework combines a benign-only unsupervised ensemble, a supervised refinement stage trained on pseudo-labelled detections, and a post hoc explainability component. In this way, UNAD+ was designed not only to detect previously unseen attacks, but also to improve the quality of these detections and provide analyst-accessible explanations for its decisions.

The experimental results on CICIDS2017 and NSL-KDD showed that the framework achieved strong overall performance. The weighted voting strategy improved the first-stage ensemble by eliminating tied votes and the benign-default bias associated with simple majority voting. At the same time, the supervised refinement stage significantly reduced false positives and improved the final detection results. In particular, the false positive rate was reduced from 11.34\% to 0.18\% on CICIDS2017 and from 6.37\% to 2.77\% on NSL-KDD. The full combined framework achieved F1-scores of 98.31\% on CICIDS2017 and 98.25\% on NSL-KDD, confirming that integrating weighted ensemble detection with supervised refinement provides a substantial improvement over the original UNAD baseline.

The analysis also showed that the benefits of the framework were not limited to aggregate performance. At the class level, the second-stage classifier significantly improved the detection of several difficult classes, including SSH-Patator and DoS Slowhttptest on CICIDS2017, and R2L and U2R on NSL-KDD. At the same time, the results highlighted an important limitation: very rare attack classes remained difficult to improve when only a small number of pseudo-labelled instances reached the supervised stage. This indicates that the effectiveness of the refinement component depends not only on classifier design, but also on the availability of sufficiently informative detections from the first stage.

A further contribution of this work lies in the integration of explainability into the IDS pipeline. The local explanation layer showed that the refinement classifier relied on meaningful traffic features in representative cases, while the global surrogate model reproduced its broader behaviour with high fidelity. As a result, the framework supports not only detection, but also interpretation, auditing, and debugging. This is particularly relevant in security-critical environments where model transparency is necessary for analyst trust and real-world use.

Overall, the findings show that UNAD+ provides a coherent framework for detecting unknown network attacks by combining unsupervised anomaly detection, supervised refinement, and explainable decision support. The framework improves detection performance, reduces false positives, and enhances transparency, making it a more usable and reliable extension of the original UNAD. Future work should focus on encrypted traffic, particularly on whether stronger flow-level, timing-based, and temporal descriptors can improve detection performance and reduce false positives when visibility into packet content is limited.

% \paragraph{Paragraph headings} Use paragraph headings as needed.
% \begin{equation}
% a^2+b^2=c^2
% \end{equation}

% % For one-column wide figures use
% \begin{figure}
% % Use the relevant command to insert your figure file.
% % For example, with the graphicx package use
%   \includegraphics{example.eps}
% % figure caption is below the figure
% \caption{Please write your figure caption here}
% \label{fig:1}       % Give a unique label
% \end{figure}
% %
% % For two-column wide figures use
% \begin{figure*}
% % Use the relevant command to insert your figure file.
% % For example, with the graphicx package use
%   \includegraphics[width=0.75\textwidth]{example.eps}
% % figure caption is below the figure
% \caption{Please write your figure caption here}
% \label{fig:2}       % Give a unique label
% \end{figure*}
% %
% % For tables use
% \begin{table}
% % table caption is above the table
% \caption{Please write your table caption here}
% \label{tab:1}       % Give a unique label
% % For LaTeX tables use
% \begin{tabular}{lll}
% \hline\noalign{\smallskip}
% first & second & third  \\
% \noalign{\smallskip}\hline\noalign{\smallskip}
% number & number & number \\
% number & number & number \\
% \noalign{\smallskip}\hline
% \end{tabular}
% \end{table}

\begin{acknowledgements}
This work was partially funded by Zukunft.Niedersachsen (ZN4365).
\end{acknowledgements}

\section*{Conflict of interest}
The authors declare that they have no conflict of interest.


% \begin{thebibliography}{}
\bibliography{refs.bib}

@inproceedings{CICIDS2017,
  author    = {Iman Sharafaldin and
               Arash Habibi Lashkari and
               Ali A. Ghorbani},
  editor    = {Paolo Mori and
               Steven Furnell and
               Olivier Camp},
  title     = {Toward Generating a New Intrusion Detection Dataset and Intrusion
               Traffic Characterization},
  booktitle = {Proceedings of the 4th International Conference on Information Systems
               Security and Privacy, {ICISSP} 2018, Funchal, Madeira - Portugal,
               January 22-24, 2018},
  pages     = {108--116},
  publisher = {SciTePress},
  year      = {2018},
  url       = {https://doi.org/10.5220/0006639801080116},
  doi       = {10.5220/0006639801080116},
  bibsource = {dblp computer science bibliography, https://dblp.org}
}

@article{liu2012isolation,
  author    = {Fei Tony Liu and
               Kai Ming Ting and
               Zhi{-}Hua Zhou},
  title     = {Isolation-Based Anomaly Detection},
  journal   = {{ACM} Trans. Knowl. Discov. Data},
  volume    = {6},
  number    = {1},
  pages     = {3:1--3:39},
  year      = {2012},
  url       = {https://doi.org/10.1145/2133360.2133363},
  doi       = {10.1145/2133360.2133363},
  bibsource = {dblp computer science bibliography, https://dblp.org}
}

@article{gunning2019darpa,
  author    = {David Gunning and
               David W. Aha},
  title     = {DARPA's Explainable Artificial Intelligence {(XAI)} Program},
  journal   = {{AI} Mag.},
  volume    = {40},
  number    = {2},
  pages     = {44--58},
  year      = {2019},
  url       = {https://doi.org/10.1609/aimag.v40i2.2850},
  doi       = {10.1609/aimag.v40i2.2850},
  bibsource = {dblp computer science bibliography, https://dblp.org}
}

@inproceedings{chawla2018host,
  author    = {Ashima Chawla and
               Brian Lee and
               Sheila Fallon and
               Paul Jacob},
  title     = {Host Based Intrusion Detection System with Combined {CNN/RNN} Model},
  booktitle = {{ECML} {PKDD} 2018 Workshops - Nemesis 2018, UrbReas 2018, SoGood
               2018, IWAISe 2018, and Green Data Mining 2018, Dublin, Ireland, September
               10-14, 2018, Proceedings},
  series    = {Lecture Notes in Computer Science},
  volume    = {11329},
  pages     = {149--158},
  publisher = {Springer},
  year      = {2018},
  url       = {https://doi.org/10.1007/978-3-030-13453-2\_12},
  doi       = {10.1007/978-3-030-13453-2\_12},
  bibsource = {dblp computer science bibliography, https://dblp.org}
}

@article{chawla2002smote,
  author    = {Nitesh V. Chawla and
               Kevin W. Bowyer and
               Lawrence O. Hall and
               W. Philip Kegelmeyer},
  title     = {{SMOTE:} Synthetic Minority Over-sampling Technique},
  journal   = {J. Artif. Intell. Res.},
  volume    = {16},
  pages     = {321--357},
  year      = {2002},
  url       = {https://doi.org/10.1613/jair.953},
  doi       = {10.1613/jair.953},
  bibsource = {dblp computer science bibliography, https://dblp.org}
}

@article{aas2021explaining,
  author    = {Kjersti Aas and
               Martin Jullum and
               Anders L{\o}land},
  title     = {Explaining individual predictions when features are dependent: More
               accurate approximations to Shapley values},
  journal   = {Artif. Intell.},
  volume    = {298},
  pages     = {103502},
  year      = {2021},
  url       = {https://doi.org/10.1016/j.artint.2021.103502},
  doi       = {10.1016/j.artint.2021.103502},
  bibsource = {dblp computer science bibliography, https://dblp.org}
}

@inproceedings{lundberg2017unified,
  author    = {Scott M. Lundberg and
               Su{-}In Lee},
  editor    = {Isabelle Guyon and
               Ulrike von Luxburg and
               Samy Bengio and
               Hanna M. Wallach and
               Rob Fergus and
               S. V. N. Vishwanathan and
               Roman Garnett},
  title     = {A Unified Approach to Interpreting Model Predictions},
  booktitle = {Advances in Neural Information Processing Systems 30: Annual Conference
               on Neural Information Processing Systems 2017, December 4-9, 2017,
               Long Beach, CA, {USA}},
  pages     = {4765--4774},
  year      = {2017},
  url       = {https://proceedings.neurips.cc/paper/2017/hash/8a20a8621978632d76c43dfd28b67767-Abstract.html},
  bibsource = {dblp computer science bibliography, https://dblp.org}
}

@inproceedings{alzubi2021towards,
  author    = {Saif Alzubi and
               Frederic T. Stahl and
               Mohamed Medhat Gaber},
  editor    = {Khalid Al{-}Begain and
               Mauro Iacono and
               Lelio Campanile and
               Andrzej Bargiela},
  title     = {Towards Intrusion Detection Of Previously Unknown Network Attacks},
  booktitle = {Proceedings of the 35th International {ECMS} International Conference
               on Modelling and Simulation, {ECMS} 2021, Virtual Event, UK, May 31
               - June 2, 2021},
  pages     = {35--41},
  publisher = {European Council for Modeling and Simulation},
  year      = {2021},
  url       = {https://doi.org/10.7148/2021-0035},
  doi       = {10.7148/2021-0035},
  bibsource = {dblp computer science bibliography, https://dblp.org}
}

@article{pinto2023survey,
  title={Survey on intrusion detection systems based on machine learning techniques for the protection of critical infrastructure},
  author={Pinto, Andrea and Herrera, Luis-Carlos and Donoso, Yezid and Gutierrez, Jairo A},
  journal={Sensors},
  volume={23},
  number={5},
  pages={2415},
  year={2023},
  publisher={MDPI}
}

@article{de2023unsupervised,
  title={Unsupervised gan-based intrusion detection system using temporal convolutional networks and self-attention},
  author={de Araujo-Filho, Paulo Freitas and Naili, Mohamed and Kaddoum, Georges and Fapi, Emmanuel Thepie and Zhu, Zhongwen},
  journal={IEEE Transactions on Network and Service Management},
  volume={20},
  number={4},
  pages={4951--4963},
  year={2023},
  publisher={IEEE}
}

@article{tong2024real,
  title={A Real-Time Label-Free Self-Supervised Deep Learning Intrusion Detection for Handling New Type and Few-Shot Attacks in IoT Networks},
  author={Tong, Jianheng and Zhang, Ying},
  journal={IEEE Internet of Things Journal},
  year={2024},
  publisher={IEEE}
}

@inproceedings{breunig2000lof,
  title={LOF: identifying density-based local outliers},
  author={Breunig, Markus M and Kriegel, Hans-Peter and Ng, Raymond T and Sander, J{\"o}rg},
  booktitle={Proceedings of the 2000 ACM SIGMOD international conference on Management of data},
  pages={93--104},
  year={2000}
}

@article{halabaku2024overfitting,
  title={Overfitting in Machine Learning: A Comparative Analysis of Decision Trees and Random Forests.},
  author={Halabaku, Erblin and Byty{\c{c}}i, Eliot},
  journal={Intelligent Automation \& Soft Computing},
  volume={39},
  number={6},
  year={2024}
}

@article{aria2021comparison,
  title={A comparison among interpretative proposals for Random Forests},
  author={Aria, Massimo and Cuccurullo, Corrado and Gnasso, Agostino},
  journal={Machine Learning with Applications},
  volume={6},
  pages={100094},
  year={2021},
  publisher={Elsevier}
}

@inproceedings{ribeiro2016should,
  title={" Why should i trust you?" Explaining the predictions of any classifier},
  author={Ribeiro, Marco Tulio and Singh, Sameer and Guestrin, Carlos},
  booktitle={Proceedings of the 22nd ACM SIGKDD international conference on knowledge discovery and data mining},
  pages={1135--1144},
  year={2016}
}

@inproceedings{herbinger2023leveraging,
  title={Leveraging Model-Based Trees as Interpretable Surrogate Models for Model Distillation},
  author={Herbinger, Julia and Dandl, Susanne and Ewald, Fiona K and Loibl, Sofia and Casalicchio, Giuseppe},
  booktitle={European Conference on Artificial Intelligence},
  pages={232--249},
  year={2023},
  organization={Springer}
}

@incollection{parhizkari2023anomaly,
  title={Anomaly detection in intrusion detection systems},
  author={Parhizkari, Siamak},
  booktitle={Anomaly Detection-Recent Advances, AI and ML Perspectives and Applications},
  year={2023},
  publisher={IntechOpen}
}

@article{hou2022handling,
  title={Handling labeled data insufficiency: Semi-supervised learning with self-training mixup decision tree for classification of network attacking traffic},
  author={Hou, Yubo and Teo, Sin G and Chen, Zhenghua and Wu, Min and Kwoh, Chee-Keong and Truong-Huu, Tram},
  journal={IEEE Transactions on Dependable and Secure Computing},
  year={2022},
  publisher={IEEE}
}

@article{nisioti2018intrusion,
  title={From intrusion detection to attacker attribution: A comprehensive survey of unsupervised methods},
  author={Nisioti, Antonia and Mylonas, Alexios and Yoo, Paul D and Katos, Vasilios},
  journal={IEEE Communications Surveys \& Tutorials},
  volume={20},
  number={4},
  pages={3369--3388},
  year={2018},
  publisher={IEEE}
}

@article{qiu2023unraveling,
  title={Unraveling false positives in unsupervised defect detection models: A study on anomaly-free training datasets},
  author={Qiu, Ji and Shi, Hongmei and Hu, Yuhen and Yu, Zujun},
  journal={Sensors},
  volume={23},
  number={23},
  pages={9360},
  year={2023},
  publisher={MDPI}
}

@article{awad2025enhanced,
  title={An enhanced ensemble defense framework for boosting adversarial robustness of intrusion detection systems},
  author={Awad, Zeinab and Zakaria, Magdy and Hassan, Rasha},
  journal={Scientific Reports},
  volume={15},
  number={1},
  pages={14177},
  year={2025},
  publisher={Nature Publishing Group UK London}
}

@article{alhowaide2021ensemble,
  title={Ensemble detection model for IoT IDS},
  author={Alhowaide, Alaa and Alsmadi, Izzat and Tang, Jian},
  journal={Internet of Things},
  volume={16},
  pages={100435},
  year={2021},
  publisher={Elsevier}
}

@article{zoppi2021prepare,
  title={Prepare for trouble and make it double! Supervised--Unsupervised stacking for anomaly-based intrusion detection},
  author={Zoppi, Tommaso and Ceccarelli, Andrea},
  journal={Journal of Network and Computer Applications},
  volume={189},
  pages={103106},
  year={2021},
  publisher={Elsevier}
}

@inproceedings{kale2022hybrid,
  title={A hybrid deep learning anomaly detection framework for intrusion detection},
  author={Kale, Rahul and Lu, Zhi and Fok, Kar Wai and Thing, Vrizlynn LL},
  booktitle={2022 IEEE 8th Intl Conference on Big Data Security on Cloud (BigDataSecurity), IEEE Intl Conference on High Performance and Smart Computing,(HPSC) and IEEE Intl Conference on Intelligent Data and Security (IDS)},
  pages={137--142},
  year={2022},
  organization={IEEE}
}

@inproceedings{tavallaee2009detailed,
  title={A detailed analysis of the KDD CUP 99 data set},
  author={Tavallaee, Mahbod and Bagheri, Ebrahim and Lu, Wei and Ghorbani, Ali A},
  booktitle={2009 IEEE symposium on computational intelligence for security and defense applications},
  pages={1--6},
  year={2009},
  organization={Ieee}
}

@article{hnamte2023novel,
  title={A novel two-stage deep learning model for network intrusion detection: LSTM-AE},
  author={Hnamte, Vanlalruata and Nhung-Nguyen, Hong and Hussain, Jamal and Hwa-Kim, Yong},
  journal={Ieee Access},
  volume={11},
  pages={37131--37148},
  year={2023},
  publisher={IEEE}
}

@inproceedings{nolle2023explanations,
  title={On Explanations for Hybrid Artificial Intelligence},
  author={Nolle, Lars and Stahl, Frederic and El-Mihoub, Tarek},
  booktitle={International Conference on Innovative Techniques and Applications of Artificial Intelligence},
  pages={3--15},
  year={2023},
  organization={Springer}
}

@article{hassija2024interpreting,
  title={Interpreting black-box models: a review on explainable artificial intelligence},
  author={Hassija, Vikas and Chamola, Vinay and Mahapatra, Atmesh and Singal, Abhinandan and Goel, Divyansh and Huang, Kaizhu and Scardapane, Simone and Spinelli, Indro and Mahmud, Mufti and Hussain, Amir},
  journal={Cognitive Computation},
  volume={16},
  number={1},
  pages={45--74},
  year={2024},
  publisher={Springer}
}

@ARTICLE{11173696,
  author={Meng, Qianwei and Tao, Jing and Yuan, Qingjun and Li, Guangsong and Wang, Yongjuan and Gao, Bing and Lu, Siqi},
  journal={IEEE Transactions on Information Forensics and Security}, 
  title={Detection of Unknown Attacks Through Encrypted Traffic: A Gaussian Prototype-Aided Variational Autoencoder Framework}, 
  year={2025},
  volume={20},
  number={},
  pages={10652-10667},
  doi={10.1109/TIFS.2025.3612141}}

@ARTICLE{11131124,
  author={Ennaji, Sabrine and de Gaspari, Fabio and Hitaj, Dorjan and Kbidi, Alicia and Vincenzo Mancini, Luigi},
  journal={IEEE Access}, 
  title={Adversarial Challenges in Network Intrusion Detection Systems: Research Insights and Future Prospects}, 
  year={2025},
  volume={13},
  number={},
  pages={148613-148645},
  doi={10.1109/ACCESS.2025.3600984}}

@ARTICLE{11153648,
  author={Shaukat, Syed Usman and Khan, Saad and Parkinson, Simon},
  journal={IEEE Access}, 
  title={A Review on Multi-Step Attack Detection}, 
  year={2025},
  volume={13},
  number={},
  pages={161779-161805},
  doi={10.1109/ACCESS.2025.3607497}}

@article{ZOHOURIAN2024104034,
title = {IoT-PRIDS: Leveraging packet representations for intrusion detection in IoT networks},
journal = {Computers \& Security},
volume = {146},
pages = {104034},
year = {2024},
issn = {0167-4048},
doi = {https://doi.org/10.1016/j.cose.2024.104034},
url = {https://www.sciencedirect.com/science/article/pii/S0167404824003390},
author = {Alireza Zohourian and Sajjad Dadkhah and Heather Molyneaux and Euclides Carlos Pinto Neto and Ali A. Ghorbani}
}

@article{ahmad2022deep,
  title={A deep learning ensemble approach to detecting unknown network attacks},
  author={Ahmad, Rasheed and Alsmadi, Izzat and Alhamdani, Wasim and Tawalbeh, Lo'ai},
  journal={Journal of Information Security and Applications},
  volume={67},
  pages={103196},
  year={2022},
  publisher={Elsevier}
}

@article{zoppi2023algorithm,
  title={Which algorithm can detect unknown attacks? Comparison of supervised, unsupervised and meta-learning algorithms for intrusion detection},
  author={Zoppi, Tommaso and Ceccarelli, Andrea and Puccetti, Tommaso and Bondavalli, Andrea},
  journal={Computers \& Security},
  volume={127},
  pages={103107},
  year={2023},
  publisher={Elsevier}
}

@article{sagi2018ensemble,
  title={Ensemble learning: A survey},
  author={Sagi, Omer and Rokach, Lior},
  journal={Wiley interdisciplinary reviews: data mining and knowledge discovery},
  volume={8},
  number={4},
  pages={e1249},
  year={2018},
  publisher={Wiley Online Library}
}

@article{ibrahim2025optimized,
  title={An optimized hybrid ensemble machine learning model combining multiple classifiers for detecting advanced persistent threats in networks},
  author={Ibrahim, Nadim and Rajalakshmi, NR and Sivakumar, V and Sharmila, L},
  journal={Journal of Big Data},
  volume={12},
  number={1},
  pages={212},
  year={2025},
  publisher={Springer}
}

@article{srivastava2025arlhnids,
  title={ARLHNIDS-IoT: An accurate and robust lightweight hybrid-NIDS for IoT network security},
  author={Srivastava, Arpita and Sinha, Ditipriya},
  journal={Computers \& Security},
  volume={156},
  pages={104515},
  year={2025},
  publisher={Elsevier}
}

@article{cheng2026ensemble,
  title={Ensemble-based detection of distributed denial-of-service attacks in IoT networks using majority decision mechanisms},
  author={Cheng, Suha and Feng, Xu},
  journal={Scientific Reports},
  year={2026},
  publisher={Nature Publishing Group UK London}
}

@article{gaspar2024explainable,
  title={Explainable AI for intrusion detection systems: LIME and SHAP applicability on multi-layer perceptron},
  author={Gaspar, Diogo and Silva, Paulo and Silva, Catarina},
  journal={IEEE Access},
  volume={12},
  pages={30164--30175},
  year={2024},
  publisher={IEEE}
}

@article{arreche2024xai,
  title={E-XAI: Evaluating black-box explainable AI frameworks for network intrusion detection},
  author={Arreche, Osvaldo and Guntur, Tanish R and Roberts, Jack W and Abdallah, Mustafa},
  journal={IEEE Access},
  volume={12},
  pages={23954--23988},
  year={2024},
  publisher={IEEE}
}

@InProceedings{Alzubi2,
author="Alzubi, Saif
and Stahl, Frederic
and Al-Khafajiy, Mohammed",
editor="Bramer, Max
and Stahl, Frederic",
title="Detect, Decide, Explain: An Intelligent Framework for Zero-Day Network Attack Detection",
booktitle="Artificial Intelligence XLII",
year="2026",
publisher="Springer Nature Switzerland",
address="Cham",
pages="3--17"
}

@article{abdi2010principal,
  title={Principal component analysis},
  author={Abdi, Herv{\'e} and Williams, Lynne J},
  journal={Wiley interdisciplinary reviews: computational statistics},
  volume={2},
  number={4},
  pages={433--459},
  year={2010},
  publisher={Wiley Online Library}
}

@article{breiman1996bagging,
  title={Bagging predictors},
  author={Breiman, Leo},
  journal={Machine learning},
  volume={24},
  number={2},
  pages={123--140},
  year={1996},
  publisher={Springer}
}

% % Format for Journal Reference
% Author, Article title, Journal, Volume, page numbers (year)
% Author, Book title, page numbers. Publisher, place (year)

% etc
% \end{thebibliography}

\end{document}